# A Cyberinfrastructure-based Approach to Real Time Water Temperature Prediction

Jounghyun Lee
Konkuk University
120 Neungdong-ro, Gwangjin-gu,
Seoul 05029, Korea
82-2-450-4144
lejohy@konkuk.ac.kr

Keun Young Lee
UBITA, Konkuk University
120 Neungdong-ro, Gwangjin-gu,
Seoul 05029, Korea
82-2-450-4144
northstar21@hanmail.net

Karpjoo Jeong
Konkuk University
120 Neungdong-ro, Gwangjin-gu,
Seoul 05029, Korea
82-2-450-3510
jeongk@konkuk.ac.kr

Meilan Jiang
Konkuk University
120 Neungdong-ro, Gwangjin-gu,
Seoul 05029, Korea
82-2-450-4144
meela@konkuk.ac.kr

Bomchul Kim
Kangwon National University
192-1 Chuncheon-si, Gangwon-do, Korea
82-33-250-8572
bkim@kangwon.ac.kr

Suntae Hwang
Kookmin University
77 Jeongneung-ro, Seongbuk-gu,
Seoul, Korea
82-2-910-4748
sthwang@kookmin.ac.kr

## ABSTRACT
The prediction of water temperature is crucial for aquatic ecosystem studies and management. In this paper, we raise challenging issues in supporting real time water temperature prediction and present a system called WT-Agabus to address those issues. The WT-Agabus system is designed to be a cyberinfrastructure and to support various prediction models in a uniform way. In addition, we present a neural network-based water temperature prediction model to use only data available online from Korea Meteorological Administration (KMA). In this paper, we also show the current prototype implementation of the WT-Agabus system to support the prediction model.

## Categories and Subject Descriptors
C.2.4 [**Computer-Communication Networks**]: Distributed Systems – *distributed applications, distributed databases*; I.5.1 [**Pattern Recognition**]: Models – *Neural nets*.

## 1. INTRODUCTION
Water temperature, one of the main variables in water quality, influence many chemical processes as well as biological conditions and behaviors [1]. The development of prediction models for water temperature is crucial for aquatic ecosystem studies and management. In environmental studies, most prediction models research efforts are focused on the development of such models and their application for environmental analyses, planning, and decision-making [2, 36]. These research efforts are not intended for real time prediction.

However, real time water temperature prediction provides scientists and environmental policy makers with new opportunities for managing water quality proactively and dealing with unexpected or abrupt environmental changes promptly.

In this paper, we present a real time water temperature prediction system. First, we raise challenging issues and propose that a cyberinfrastructure should be an effective approach to addressing those issues. Second, we introduce a prediction model for water temperature that can be supported in a real time manner. Third, we present a system called WT-Agabus for real time water prediction that is designed as a cyberinfrastructure system. Cyberinfrastructure integrates computing systems, data storages, instruments, data visualization and other information services into a distributed system environment to support advanced research activities effectively in scientific and engineering domains.

Agabus is our ongoing research effort to develop a stream processing-based real time prediction system for scientific and engineering applications. WT-Agabus is a part of the development effort aimed at real time water temperature prediction. Our development strategy for Agabus is not to implement the system from scratch but to integrate available system middleware as much as possible. This strategy allows us to reduce implementation work significantly and to design the system to be composable and extensible.

In the development of WT-Agabus, we use the sensor network middleware called CSN (Conceptually Manageable Sensor network) and the cloud sensor data management system called S4EM (Simple Sensor Data Stream Management System for Environmental Monitoring) which are already developed as separate projects [3, 4]. In addition, we have newly developed a prediction model simulation system based the stream processing model (PS3) for WT-Agabus. In this paper, we introduce CSN and S4EM, explain the design and implementation of PS3, and discuss how such system middleware is integrated into a cyberinfrastructure. [5]

This paper is organized as follows. In Section 2, we raise challenging issues and propose the design approach based on cyberinfrastructure. Section 3 gives a water temperature prediction model based on neural networks. The system design is explained in Section 4. Section 5 explains the system implementation and shows the analyses on the neural network based water temperature prediction model. In Section 6 and 7, we discuss related work and conclude this paper.

## 2. CHALLENGING ISSUES AND CYBERINFRASTRUCTURE-BASED APPROACH
*Real time water temperature prediction* is the real time computational activity to collect input data (e.g., meteorological or hydrologic data at the current time), to execute prediction algorithms (i.e., models) with those data, and to deliver prediction results (i.e., water temperature data) to human users or other computer systems that decide or take actions based on prediction results. Therefore, the real time water temperature prediction requires both computing algorithms and computing systems. [6, 7]

Real time water temperature prediction raises two challenging issues:

- *Development of Real Time-Executable Prediction Models*. Prediction models must be developed to use input variables whose values can be automatically collected in a real time manner.

- *Real Time Integration of Monitoring, Prediction Model Simulation and Notification*. There are lots of research work on both environmental monitoring and the development of prediction models. However, they are usually separate studies and there is little research work on their real time integration.

In this paper, we propose a *cyberinfrastructure approach* to real time execution support for water temperature prediction models mentioned in Section 2 and 3. In this approach, the real time execution support system is designed as a cyberinfrastructure that facilitates not only *the execution support for various water temperature prediction models in a real time manner* but also *the development or improvement of such prediction models.*

Cyberinfrastructure is a well-known technological terminology, but at the same time, not a clearly defined technology [8]. In this paper, we define cyberinfrastructure:

- To integrate computing systems, data management systems (e.g., data repositories), instruments, and user interface systems (e.g., data portals and visualization) into a single but loosely coupled distributed system or environment to provide a variety of services focused on a specific application domain.

- To support administration, maintenance, and system upgrades in order to allow scientists to focus on environmental research by the facilitation of operating or maintaining computer systems.

In this paper, we argue that the effective support for real time water temperature prediction require the following issues to be addressed *in a real time and integrated way*:

- *Monitoring*. Real time environmental monitoring is required both for the development of prediction models and for the simulation of prediction models. Prediction models (e.g., empirical models such as neural network models) are developed by monitoring sites and analyzing data from monitoring. In addition, prediction models also use real time monitoring data as input data to compute prediction values.

- *Simulation*. The effective runtime execution (i.e., simulation) support for various prediction models is required. There are a variety of water temperature prediction models possible and in fact available because there are various prediction model development techniques and various environmental variables can be considered for prediction. Those prediction models require not only different execution methods or system tools but also data values for a different set of environmental variables.

- *Event Detection and Notification*. Real time prediction is usually intended to allow humans or computer systems to take preventive actions or to be prepared against future events or phenomena *in a proactive way*. Such proactive efforts require the types of events or phenomena to be defined and detected, based on prediction values. In addition, humans or computer systems for proactive actions must also be notified of the detection of events or phenomena.

- *Data Management*. Monitoring, simulation (i.e., prediction) and event detection generate data. Data from monitoring is crucial for the development, evaluation and improvement of prediction models.

- *Prediction Model Development*. There are a variety of approaches to water temperature prediction models. It is important to support the development of various models.

Figure 1 illustrates how these issues are integrated. Those arrows represent data flows.

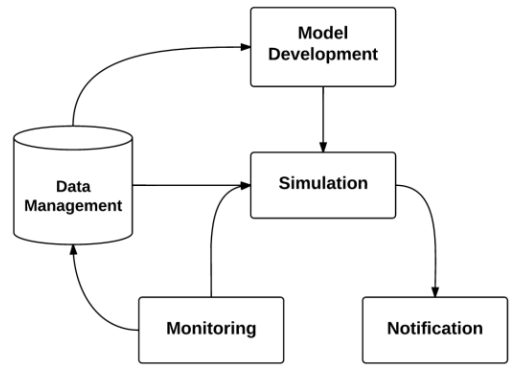

**Figure 1. The Computing Structure of Real Time Water Temperature Prediction**

## 3. NEURAL NETWORK-BASED WATER TEMPERATURE PREDICTION

In the WT-Agabus development project for real time water temperature prediction, we are currently developing a number of water temperature prediction models. These models differ in how accurate and stable their prediction results are and what input variables are required for prediction.

Real time prediction requires values for those input variables to be collected in a real time manner. Therefore, if a prediction model uses input variables whose values cannot be collected in a real time manner, then the model is not suitable for real time prediction. In this section, we introduce a prediction model for water temperature whose input variables can be easily obtained. In fact, the model choose input variables whose values are available online by Korea Meteorological Administration (KMA).

### 3.1 Sampling

Water temperature data were obtained from Soyang River basin located in a mountainous district in South Korea. The watershed is composed of 87% deciduous forest and 6% agricultural field. The length of the Soyang River is 77km, and the altitude difference of the watershed is from 200m to 1,700m. Therefore this river is a steep river system. The ratio of water flow rate on dry days and rainy days are greater than 300 in Korea due to the effect of summer monsoon. On dry days, water flow is slow and the temperature is well equilibrated with air temperature. But on rainy days, the water is cooler than air temperature, because water flows fast and it does not have enough time to equilibrate with air temperature. Therefore the cooling effect of high altitude air remains to the downstream reaches.

The water temperature used in this study was measured between the Soyang River and Lake Soyang. Water temperature was measured hourly at different intervals during 1991 and 2009. The water temperature were recorded at depth of 1 m due to the fast water flow.

Meteorological data for the river were obtained from the Inje Meteorological station, Korea during 1991-2009. The meteorological station was located within the Soyang River, less than 3 km from the Soyang River water temperature site. We used hourly estimates of air temperature (TM) and rainfall (RF) data.

## 3.2 Model development

### 3.2.1 Data preprocessing

It is well known that meteorological variables such as air temperature, solar radiation, and rainfall, influence water temperature [9, 10]. In this study, we used meteorological data including air temperature and rainfall. The measured water temperatures were considered as an output variable in our model. The data set for the entire 19 year consisting of 1126 records for all variables was used in this study. The first 790 observations were used to construct the prediction model. The remaining observations were used as test data to verify the prediction model. Artificial Neural Network (ANN) model was employed to predict water temperature in Soyang River.

To generate the input variables, we investigated the correlations coefficients with an imposed time lag. We assumed that hourly water temperature in hour $t$ ($Y_t$) was related to the values of the meteorological variables in hour $t$-$k$ ($X_{t-k}$). The association between the values of $Y_t$ and $X_{t-k}$ was measured by comparing cross-correlation between $Y_t$ and $X_{t-k}$ when $k$ varies from 0 to 24. The cross-correlation that is based on the Pearson correlation function was calculated by [11, 12].

$$\hat{\rho}_{YX}(k) = \begin{cases} \dfrac{1}{N}\dfrac{\sum_{t=1}^{N-k}(Y_t - \bar{Y})(X_{t+k} - \bar{X})}{s_Y s_X} & \text{if } k \geq 0 \\ \dfrac{1}{N}\dfrac{\sum_{t=1-k}^{N}(Y_t - \bar{Y})(X_{t+k} - \bar{X})}{s_Y s_X} & \text{if } k < 0 \end{cases}$$

Where $s_Y$ and $s_X$ are sample standard deviation of the time series $Y_t$ (mosquitoe abundances) and $X_t$ (weather variables), respectively. We applied the cross-correlation to calculate the time lag of between the air temperature and the water temperature. It is 17 hours. So, the air temperature (TM), the air temperature for the preceding 17 hours (TM$_{t-17}$), and the rainfall (RF) were selected as input variables.

To estimate prediction model performance, the root-mean-square error (RMSE), the Nash Coefficient (NASH), the correlation coefficient (R), and the index of agreement (IA) will be used, which are given by: [10, 13, 14, and 15]

$$\text{RMSE} = \sqrt{\dfrac{\sum_{i=1}^{N}(O_i - P_i)^2}{N}}$$ where $N$ is the number of hourly water temperature observations; $O_i$ the observed hourly water temperatures; $P_i$ the predicted hourly water temperatures and

$$\text{Nash} = 1 - \dfrac{\sum_{i=1}^{N}(O_i - P_i)^2}{\sum_{i=1}^{N}(O_i - \bar{O})^2}$$ where $\bar{O}$ represents the mean hourly water temperatures for the period $N$.

$$R = \sqrt{\dfrac{\sum_{i=1}^{n}(O_i - \bar{O})^2 - \sum_{i=1}^{n}(O_i - P_i)^2}{\sum_{i=1}^{n}(O_i - \bar{O})^2}}$$

$$\text{IA} = 1 - \dfrac{\sum_{i=1}^{n}|P_i - O_i|^2}{\sum_{i=1}^{n}(|P_i - \bar{O}| + |O_i - \bar{O}|)^2}$$

The IA is a relative measure which is suitable to evaluate different models to be compared using different data set [15].

### 3.2.2 Artificial Neural Network

A multi-layer ANN uses an approach that creates models of a system state using non-linear combinations of the input variables [16, 17, and 18]. The ANN applied in this paper is a feed-forward network with sigmoid functions in the hidden layers and a linear activation function in the output node in MATLAB ver. 2012a. According Bishop's research (1995), it is enough to use only one hidden layer. The ANN is trained using a backpropagation algorithm that is essentially a gradient descent technique that minimizes the network error function [19].

The ANN requires the learning rate, number of nodes in a single hidden layer, and maximum number of training epochs are specified. In this paper, we applied the optimal number error approach [20]. The number of nodes in the hidden layer was varied between 5 and 25 and the learning rate was varied from 0.01 to 1.0 in increments of 0.05. For each configuration the mean square error (MSE) between the model output and the measured data was computed. Having 10 nodes in the hidden layer and 0.55 learning rate resulted in the maximum model performance indicated by MSE. The final ANN structure had 3 input variables with one node accounting for bias, 10 hidden neurons with one node accounting for bias, a 0.55 learning rate, and one output variable of the output layer in Figure 2.

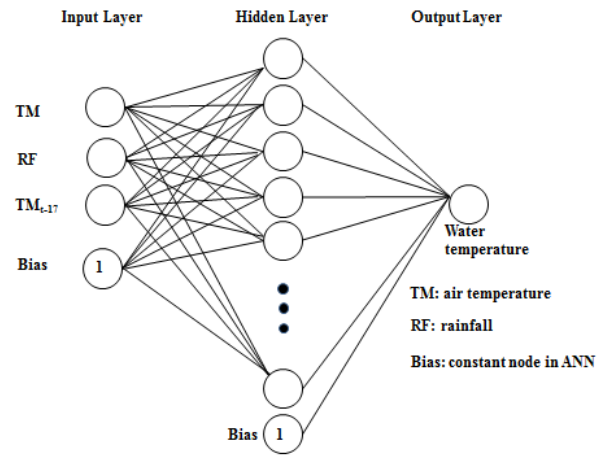

**Figure 2. A Presentation of Feedforward Artificial Neural Network**

## 4. SYSTEM DESIGN

### 4.1 System Architecture

We now present a real time water temperature prediction system called WT-Agabus. WT-Agabus is not designed for a particular prediction model, but to support various prediction models by customization. The architecture of the WT-Agabus system is presented in Figure 3 and its major system components are:

- *Stream Processing-based Prediction Model Simulation System (PS3)*. The design of PS3 is not dedicated to a particular prediction model, but designed to support various prediction models.
- *Sensor Network-based Monitoring (SNM)*. SNM is developed on a sensor network system (called CSN). CSN is a general-purpose sensor network system to facilitate the conceptual management of sensor networks and the easy application development. [4]
- *Cloud Data Repository (CDR)*. CDR is developed on a cloud NoSQL (Not only SQL) data repository (called S4EM) and manages both monitoring data and prediction results. S4EM is a Software as a Service (SaaS) system to support sensor data management.[3, 37, 38]
- *CEP-based Notification System*. WT-Agabus uses a Complex Event Processing (CEP) engine (specifically, the Esper system) to support event-driven notification. The notification system allows the user to define patterns or detection conditions about streams of prediction values. Then, the notification monitors the stream of prediction values to detect if those patterns or conditions occur in the stream.
- *Web Data Portal* (WDP). WDP provides web-based access to monitoring data and prediction results. WDP is intended to help scientists with the development and improvement of prediction models. [21]

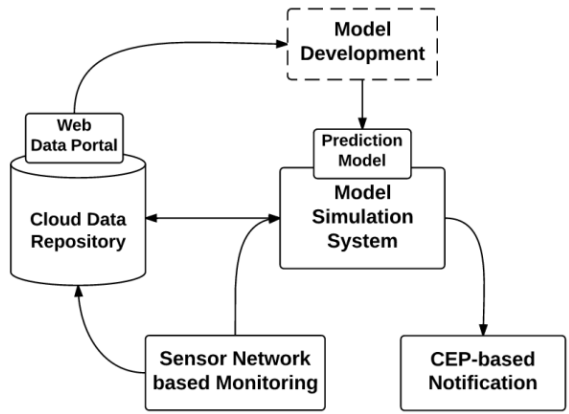

**Figure 3. Conceptual System Model for Real time prediction**

## 4.2 Monitoring, Data Management and Notification

In the development of the WT-Agabus system, we use existing system middleware (in fact, developed as separate projects) for real time monitoring, data management, and notification:

- CSN (Conceptually Manageable Sensor Network) for real time monitoring.
- S4EM (Simple Sensor Data Stream Management System for Environmental Monitoring) for data management.
- Esper for CEP-based Notification. Esper is an open source Complex Event Processing engine.

All of these middleware systems are based on stream processing and management. Therefore, they are integrated according to the stream processing model [5, 22]. In this section, we briefly introduce CSN and S4EM. Since we are currently using basic and simple CEP features in Esper, we do not explain Esper in this paper. However, we plan to develop a notification system on top of Esper [23].

### 4.2.1 CSN: Conceptually Manageable Sensor Network

In WT-Agabus, we use CSN to develop the real time monitoring system [4]. CSN is our separate project for the development of a general-purpose sensor network middleware. It is designed to facilitate the conceptual management of sensor networks and the easy application development.

In CSN, each sensor is managed as a logical data stream and implemented as a message queue. Sensors are accessed and shared via message queues *according to the publish/subscribe model*. Sensors simply send their data out by publishing them to their message queues. Applications can access data from sensors by subscribing their message queues. This design makes applications decoupled from sensors and greatly facilitates the development of applications.

In addition to the publish/subscribe model based communication, the CSN runtime system also supports a simple TCP/IP socket based communication. Sensors can simply send their data to the socket and then the CSN runtime system publishes their data for the behalf of those sensors. This communication method is intended for those sensors that cannot communicate according to the publish/subscribe model.

The current design the CSN system also supports other computer systems as a kind of virtual sensor as long as those systems are considered to generate data streams.

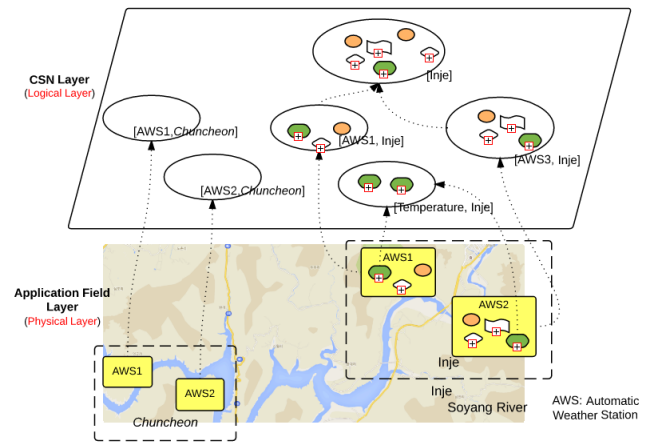

**Figure 4. Conceptual System Model for CSN**

### 4.2.2 S4EM: Simple Sensor Data Stream Management System for Environmental Monitoring

S4EM is designed to manage data streams from sensors [3]. In S4EM, a *sensor data stream* is simply a sequence of data values from a sensor and can be appended with new values until the data stream is explicitly closed. Therefore, the S4EM-based data repository consists of only a number of sensor data streams. S4EM currently assumes the type of data from a sensor to be numeric and the data is associated with timestamps. We plan to support data to be of other types. Therefore, a sensor data stream in S4EM is an infinite sequence of numeric data values.

S4EM is currently implemented as a Software as a Service (SaaS) cloud service on top of the Google Datastore, it provides a number of sensor data stream management services: create, append, search, retrieve, delete, and download. These services can be invoked by Web Browsers because they are implemented as Servlet programs. [37]

Since both CSN and S4EM assume data streams as the primary data object, they can be easily integrated according to the stream processing model [5, 22].

In addition to sensor data streams, S4EM also support the *management of metadata* about those streams. Since S4EM is intended for environmental monitoring, S4EM is designed to provide a predefined metadata model. The metadata model is designed to facilitate the analysis of sensor data. The current data model is based on the VEGA model developed by the Global Lake Ecological Observatory Network (GLEON) [24, 25, and 26].

Figure 5 shows the data model in the Entity-Relationship (ER) diagram that S4EM currently supports:

- *Stream*, *Value*, and *Sensor*. For each sensor data stream, a single data record (called a Stream object) in the *Stream* table is created to represent the stream. Each Stream object is associated with a single infinite sequence of data values in the *Value* table. Each Stream object is also linked to a sensor that generates the data stream.

- *Variable* and *Unit*. A sensor is assumed to measure the value of a property about the environment. The property is called a variable and the information about variables is maintained in the *Variable* table. The Unit table contains information units for data values.

- *Site* and *Source*. For a data stream, the information about its site is maintained in the *Site* table. The site is the place where the sensor operates. The *Source* table contains the information about who is responsible for the sensor and the data stream.

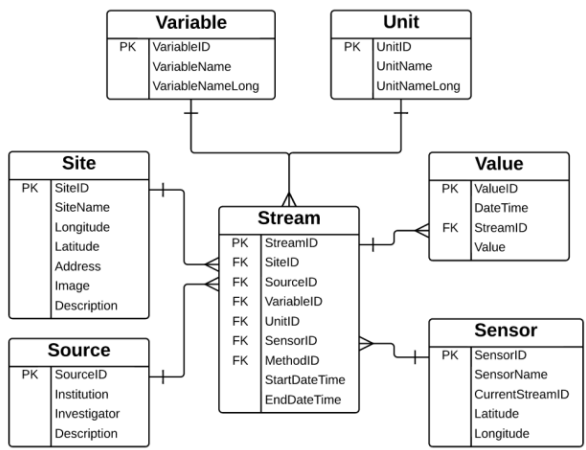

**Figure 5. ER Diagram for S4EM**

## 4.3 PS3: Stream Processing-based Prediction Model Simulation System

### 4.3.1 Design Strategy and System Architecture

In this subsection, we explain the design strategy and system architecture for the PS3 system. The main design strategy for the PS3 system is to support:

- *Heterogeneous Prediction Models*. There are a number of prediction models based on various methods such as Neural Networks [27], Hidden Markov Models [28], and Genetic Algorithms [29]. These models are implemented by model development tools such as MATLAB or R. These models can be manually programmed. PS3 intends to support them in a uniform and configurable way.

- *Heterogeneous Simulation Tools*. There are a number of simulation tools to run prediction models. They include MATLAB [30] and R [31]. PS3 intends to support them in a uniform and configurable way.

- *Stream Processing-based Continuous Simulation of Prediction Models*. PS3 runs a prediction model according to the stream processing model [32]. In this model, PS3 assumes data streaming from monitoring systems (e.g., a CSN system) and iterates the simulation of the model with data values in data streams. PS3 also sends prediction results out as data streams to a notification system or other applications. Figure 6 illustrates the stream processing-based simulation.

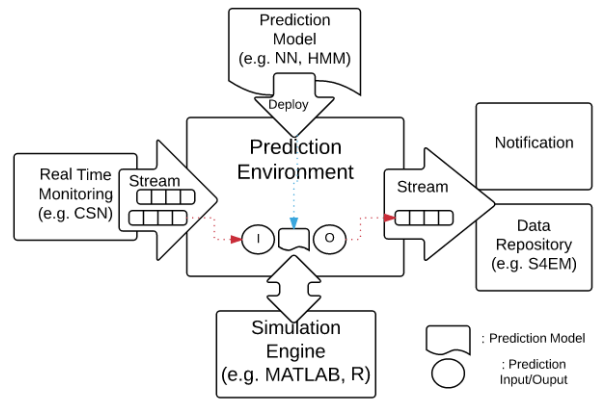

**Figure 6. Stream Processing-based Continuous Simulation of Prediction Models**

The system architecture of the PS3 system is shown in Figure 7 and major system components are as follows:

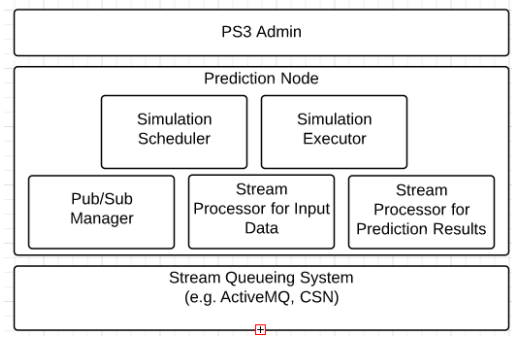

**Figure 7. PS3 System Architecture**

- *Pub/Sub Manager*. PS3 and the other systems (CSN and S4EM) are connected via external data streams implemented as message queues. In the PS3 system, the Pub/Sub (Publish/Subscribe) Manager connects to those external data streams according to the publish/subscribe model. In the current PS3 system, the Pub/Sub Manager subscribes the data streams (monitoring data) to be published by CSN and publishes the data streams (prediction results) to be subscribed by the notification system and other prediction applications.
- *Simulation Scheduler*. PS3 runs prediction models according to a scheduling mode: On-demand Prediction, Time-scheduled Prediction, and Data-driven Prediction. The Simulation Scheduler manages the schedule. According to the schedule, it runs prediction models by invoking the Stream Processors for input data, the Simulation Executor, and the Stream Processor for prediction results.
- *Stream Processor for Input Data (SPID)*. SPID processes and reorganizes data in data streams to generate input data required for prediction models. SPID receives data stream from the Pub/Sub Manager.
- *Simulation Executor (SE)*. SE starts the simulation engine (e.g., MATLAB or R) with the prediction model specified in the current configuration file.
- *Stream Processor for Prediction Results (SPPR)*. SPPR parses and send prediction results from the simulation of the prediction model to the Pub/Sub Manager.
- *PS3 Admin*. PS Admin provides administration operations, data stream management, and prediction model management as RESTful Web Services. Therefore, Web applications can invoke those admin services.

Figure 8 illustrates the runtime structure of PS3.

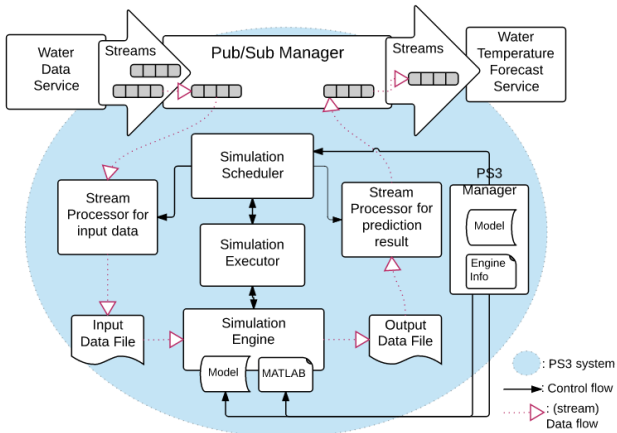

**Figure 8. PS3 runtime system**

### 4.3.2 Simulation Scheduler

PS3 does not execute the simulation of a prediction model for each data value in data streams. PS3 runs the simulation according to a schedule that is independent of the arrival of data values. For example, PS3 may run the simulation at every hour although data values arrive at every minute. PS3 supports the following four schedule modes:

A. *On-demand Prediction*. The simulation of a prediction model is explicitly requested by the user or external systems.

B. *Time-Scheduled Prediction*. The time schedule for prediction is registered. A specific time can be registered for one point prediction. In addition, the list of a start time, a time interval, and the number of simulation executions or an end time can be given for multiple time prediction.

C. *Event-driven Prediction*. Event patterns over data streams are registered and the simulation is run on the detection of a pattern.

D. *Data-driven Prediction*. On every arrival of data values, the simulation is started.

When it decides to run the simulation of a prediction model, the Simulation Scheduler invokes the Stream Processor for Input Data (SPID), the Simulation Executor, and the Stream Processor for Prediction Results.

### 4.3.3 Stream Processors for Input Data and Prediction Results

Prediction models often require computational processing or the selection of a specific part of a data stream for input data. For example, the prediction model given in Section 3 requires two input data: the average of air temperature during the last one hour and the average of air temperature during one hour before 17 hours. Such input data cannot be generated with stream processing inside PS3.

PS3 supports the Stream Processor for Input Data (SPID) to deal with such input data requests to require stream processing. In PS3, SPID uses the Esper complex event processing Engine to support stream processing because complex event processing is similar to stream processing and Esper provides rich functionality. Esper supports the event programming language (EPL) to specify event processing operations and such EPL programs are called EPL statements. PS3 allows the user to register EPL statements as input data generation rules to be used by SPID. [33, 34]

The design of the Stream Processor for Prediction Results (SPPR) is the same as that of SPID. However, since no stream processing is usually required for prediction results, the current implementation of SPPR simply publishes prediction results as the prediction model produces

Figure 9 illustrates how the Stream Processor for Input Data (SPID) works. SPID receives data values in data streams in CSN and construct local data streams in Esper (in fact, event streams in Esper). Esper runs registered EPL statements to generate input data (e.g., the average of data values for the last one hour). Those input data values are currently saved as an input data file.

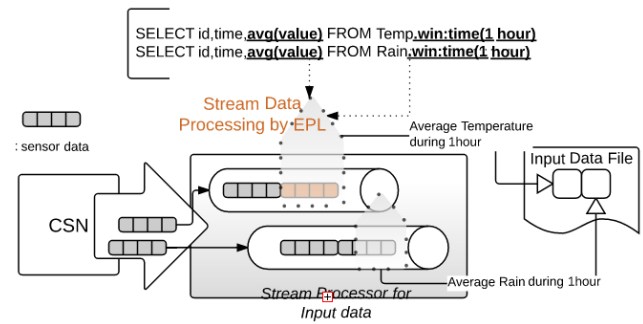

**Figure 9. Stream Processor for Input Data**

### 4.3.4 Simulation Executor

PS3 is designed to support various prediction models and simulation tools. For such heterogeneity support, PS3 provides a method to specify prediction models and simulation tools in a standardized way. In order to manage various prediction models and simulation tools in a uniform way, PS3 define Prediction Model Archive (PMA) that is standard format of prediction model. PMA consists of configuration files, script files, input data sets and their directory organization shown in Figure 10. PMA is actually compressed file in PS3.

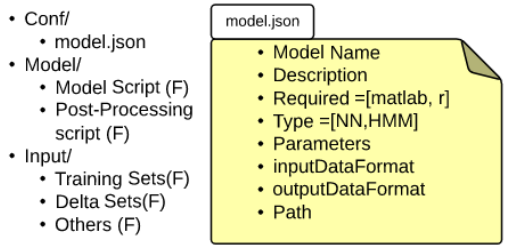

Figure 10. Structure for Prediction Model Archive (PMA)

## 5. IMPLEMENTATION AND EXPERIMENTS

### 5.1 The current prototype implementation of WT-Agabus

In this section, we explain the current prototype implementation of the WT-Agabus system. The implementation is shown in Figure 11. The WT-Agabus prototype runs the prediction model in Section 3 and receives input data from KMA (Korea Meteorological Administration) [35].

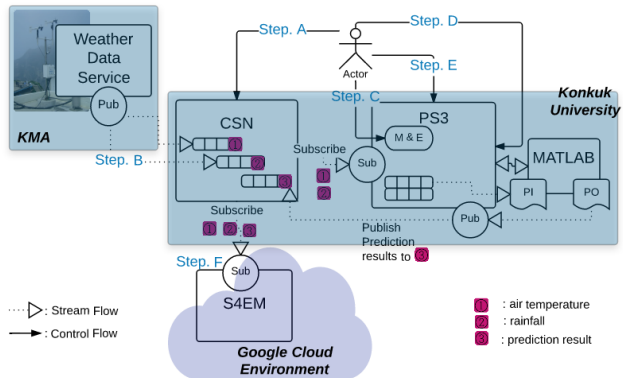

Figure 11. Runtime Structure of the Current Prototype of WT-Agabus

A. *Register three sensors and create three data streams for those sensors in the CSN system.* Currently, two sensors from Korea Meteorological Administration (KMA) are air temperature and rainfall. These sensors are numbered 1 and 2 in the Figure 11. The other sensor numbered 3 in the Figure 11 represents prediction results (i.e., water temperature) from PS3. In this implementation, we assumed virtual sensors for prediction results in such way that prediction models are considered as those virtual sensors.

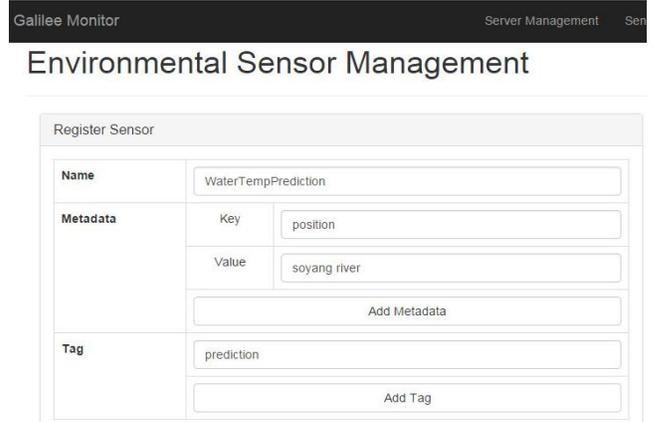

Figure 12. Sensor Registration Window in CSN

B. *Read data from two KMA sensors and publish them to the CSN system.* KMA does not allow other systems to access their sensors directly, but provide RESTful web services (http://newsky2.kma.go.kr/service/SecndSrtpdFrcstInfoService/ForecastGrib) for access to data from those sensors. The RESTful interface is shown in Table 1 and Table 2. Currently, the KMA provides data about ten weather parameters such as temperature, rainfall, wind speed, wind direction. We implemented a process to read data from sensors via the RESTful web services and to publish data to CSN.

Table 1. Request Parameters for Weather Data Service

| Parameter | Sample Data | Introduction |
|---|---|---|
| base_date | 20121206 | The rises. |
| base_time | 1100 | The time |
| Nx | 1 | The spot X coordinate (relative position of latitude) |
| Ny | 1 | The spot Y coordinate (relative position of longitude) |

Table 2. Response Parameters for Weather Data Service

| Parameter | Sample Data | Introduction |
|---|---|---|
| resultCode | 0 | Result-code |
| resultMsg | OK | The result message error message |
| numOfRows | 10 | The number of Rows |
| pageNo | 1 | Page number |
| totalCount | 10 | The number of aggregated result |
| category | LGT | Data division code |
| obsrValue | -1 | The real condition value |

C. *Register the prediction model (given in Section 3) and a simulation tool (MATLAB) in the PS3 system.* The information about the prediction model and the simulation tool is given to PS3 as a PMA file whose format is explained in Section 4.3). PS3 provides the RESTful interface for the registration of prediction models and simulation tools. The interface is given in Table 3.

**Table 3. Interface for Managing Prediction Model and Simulation Tool**

| # | Action | Method | Resources |
|---|---|---|---|
| 1 | Register the prediction model | PUT | /node/models |
| 2 | Get the prediction model | GET | /node/models/[mid] |
| 3 | Remove the prediction model | DELETE | /node/models/[mid] |
| 4 | Register the simulation tool | PUT | /node/engines |
| 5 | Get the information of simulation tool | GET | /node/engines/[eid] |
| 6 | Remove the simulation tool | DELETE | /node/engines/[eid] |

D. *Register three data streams (created in Step A) and stream processing rules (i.e., EPL statements) for input data in the PS3 system.* Data streams are registered in PS3. PS3 reads data from two data streams in CSN by subscribing their data streams. PS3 publish prediction results to the other data stream in CSN. The Stream Processor for Input Data (SPID) in PS3 generates input data from data streams in CSN by using the rules. Specifically, SPID generates three input variables: the average of air temperature during one hour, the average of air temperature during one hour before 17 hours, and the average of rainfall during one hour. Table 4 shows the EPL rules for input data generation.

**Table 4. EPL Rules for Input Data Generation**

| Stream in CSN | EPL Statements | Idx |
|---|---|---|
| air temperature | SELECT id, timestamp, avg(value) FROM Temperature.win:time(1 hour) | 0 |
| | SELECT id, timestamp, avg(value) FROM Temperature.win:time(17 hour) output snapshot every 1 hour | 2 |
| rainfall | SELECT id, timestamp, avg(value) FROM Rainfall.win:time(1 hour) | 1 |

E. *Set the prediction schedule mode and start prediction.* The prediction schedule mode is currently set to the time scheduled mode. Table 5 shows interface for managing prediction.

**Table 5. Interface for Managing Prediction**

| # | Action | Method | Resources |
|---|---|---|---|
| 1 | Register Prediction Mode<br>[mode]: prediction type<br>1:On_demand, 2: Scheduled. 3:Data_driven<br>[interval]: interval time to re-run prediction model | POST | /node/prediction?mode=1&time=t&interval=3600 |
| 2 | Start Prediction | POST | /node/prediction?action=start |
| 3 | Stop Prediction | POST | /node/prediction?action=stop |

F. *Subscribe three data streams (created in Step A) in CSN and store prediction results in S4EM.* Figure 13 shows the snapshots of the S4EM windows.

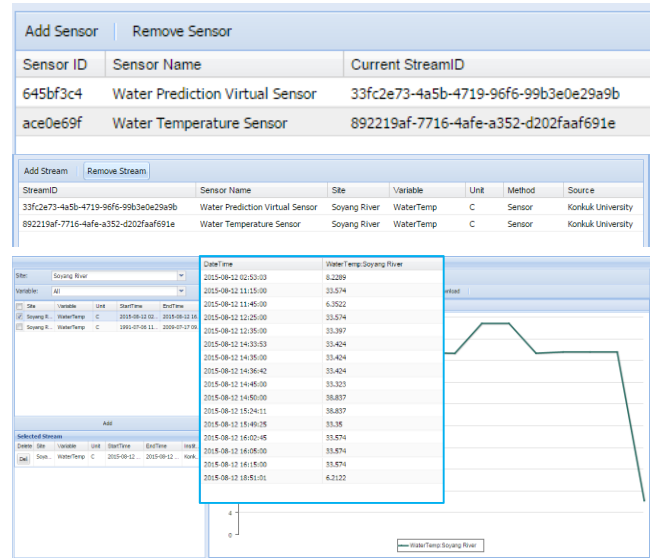

**Figure 13. (a) Sensor Management Window, (b) Data Stream Management Window, (c) Search and Download Window for Prediction Results**

## 5.2 Evaluation of the Water Temperature Prediction Model

WT-Agabus with ANN was applied to forecast the water temperature of the Soyang River. The meteorological properties and the water temperature at the test site are provided in Table 6. High variability was observed in both RF and WT.

**Table 6. Mean and Standard Deviation Values of Meteorological Characteristics and Water Temperatures during the 19 Experimental Periods**

| | *TM. (℃)* | *RF. (mm)* | *WT. (℃)* |
|---|---|---|---|
| Mean±std. | 17.26±7.65 | 1.64±4.92 | 16.89±5.83 |
| Ranges | -11.5~30.7 | 0~66 | -12.3~28.8 |

Results showed that the ANN model performed well. The ANN provides both of relatively good IA value of 0.92 and R value of 0.81 for the model with RMSE of 3.15 and NASH of 0.64 (Table C). Figure 14 shows the performance of the ANN for predicting test dataset. In Figure 14, the sample number is used for 336 water temperature records in the X axis instead of time in order to show difference between measured and predicted data clearly. This shows that the ANN model predicted well for low and average values, while it could not predict relatively high values. The ANN model tried to catch the fluctuation patterns in the water temperatures. However, water temperatures were often overestimated.

**Table 7. ANN model performance statistics**

|  | Mean | STD | RMSE | NASH | IA | R |
|---|---|---|---|---|---|---|
| ANN | 16.91 | 5.22 | 3.15 | 0.64 | 0.92 | 0.81 |

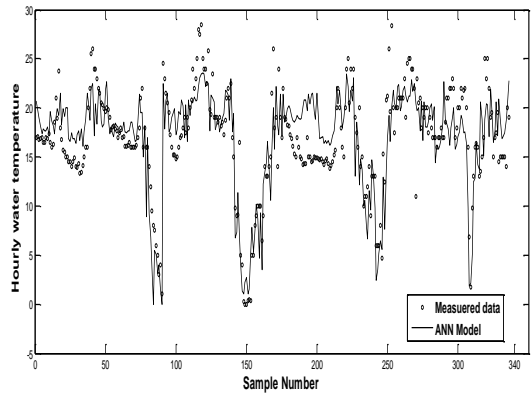

**Figure 14. Measured and Predicted Water Temperatures in the Soyang River Obtained by ANN**

The performance of the ANN model could be improved if more variables are added. However, since we can access only two real time meteorological data, i.e. air temperature and rainfall in an online manner, we conducted the experiments with this limited set of variables.

## 6. RELATED WORK

There are a variety of approaches such as statistical models [41, 42, 43, and 44] and deterministic models [9, 45, and 46] to predict water temperature. These research approaches are not intended for real time prediction. This paper is aimed at the real time support for the prediction of water temperature.

There are lots of research work on sensor networks and cloud databases [24, 25, 37, 38, and 39]. This paper addresses issues in integrating those technologies into a cyberinfrastructure for real time prediction.

There are active ongoing research activities on distributed streaming processing systems [22, 23, 32, 33, and 34]. These systems are focused on individual data object-level processing. This paper addresses issues in integrating streaming processing into complicated computations such as prediction.

## 7. CONCLUSIONS AND FUTURE WORK

In this paper, we presented the real time water temperature prediction system called WT-Agabus. The prediction of water temperature is crucial for aquatic ecosystem studies and management. We raised challenging issues in real time water temperature prediction and explained how these issues are addressed by the cyberinfrastructure-based design.

In the cyberinfrastructure-based design, we integrate sensor networks (CSN), cloud databases (S4EM), the prediction model simulation system (PS3) and the Esper complex event processing system in a real time manner. The integration is based on the distributed stream processing model [32, 33, and 34].

In addition, we also present a neural network-based prediction model for water temperature that uses only data available online from Korea Administration Agency. In this paper, we also showed the current prototype implementation of the WT-Agabus system to support the prediction model.

## 8. ACKNOWLEDGMENTS
This work was funded by the Weather Information Service Engine Project under Grant KMIPA-2012-0001.